\documentclass[prl,twocolumn,showpacs,superscriptaddress,floatfix,letter]{revtex4}

\usepackage[dvips]{graphicx}    
\usepackage{verbatim}           
\usepackage{amsmath}            

 \newcommand{\etal}{\textit{et~al}.}

 \renewcommand{\comment}[1]{}
 \renewcommand\it{\textit}
 \renewcommand\bf{\textbf}

 \newcommand{\cm}{cm$^{-1}$}

\begin{document}


\title{Length-dependent Raman spectroscopy of single-walled carbon
nanotubes: \\ the effect of dispersant on defects }

\comment{ Quantifying defects in length-separated single-walled
carbon nanotubes with Raman spectroscopy: the effect of
dispersant}


\author{J.~R.~Simpson}
\email[]{jrsimpson@towson.edu}
 \affiliation{ Department of Physics, Astronomy, and Geosciences,
 Towson University, Towson, MD 21252, USA.}
 \affiliation{ Physics Laboratory,
National Institute of Standards and Technology, Gaithersburg, MD
20899, USA.}
\author{J.~A.~Fagan}
\affiliation{ Materials Science and Engineering Laboratory,
National Institute of Standards and Technology, Gaithersburg, MD
20899 USA.}
\author{M.~L.~Becker}
\affiliation{ Materials Science and Engineering Laboratory,
National Institute of Standards and Technology, Gaithersburg, MD
20899 USA.}
\author{E.~K.~Hobbie}
\affiliation{ Materials Science and Engineering Laboratory,
National Institute of Standards and Technology, Gaithersburg, MD
20899 USA.}
\author{A.~R.~Hight~Walker}
\affiliation{ Physics Laboratory, National Institute of Standards
and Technology, Gaithersburg, MD 20899, USA.}



\date{November 2, 2008}

\begin{abstract}
We compare Raman spectra from aqueous suspensions of
length-separated single-walled carbon nanotubes (SWCNTs) dispersed
using either polymer adsorption of single-stranded DNA or miscelle
encapsulation with sodium deoxycholate surfactant.  The Raman
spectral features, other than the D-band, increase monotonically
with nanotube length in both dispersion schemes.  The intensity
ratio of the disorder-induced D to G$'$ Raman bands decays as a
function of SWCNT length, proportional to $1/L$, as expected for
endcap defects. While the UV-vis absorption and fluorescence also
increase with length for both dispersants, the fluorescence
intensity is dramatically lower for DNA-wrapped SWCNTs of equal
length.  The similarities in the length-dependent D/G$'$ ratios
exclude defects as an explanation for the fluorescence decrease in
DNA versus deoxycholate dispersions.

\comment{}

\end{abstract}


\maketitle


Applications of single-walled carbon nanotubes (SWCNTs) require an
understanding of their intrinsic characteristics and sample
quality. Recently, highly-purified SWCNTs have been produced that
show substantial property improvement over unsorted
materials.\cite{kitiyanan03} However, many of the separation
schemes, whether for electronic type or length, rely on different
agents to aid in the dispersion of individual SWCNTs prior to
separation.  In this letter, we address the possible effect of the
dispersing agent on the fundamental SWCNT optical characteristics,
in particular the Raman scattering, either through interaction
with the nanotube structure or from inherent processing
differences. We accomplish this by comparing length-separated
fractions of the same nanotube batch dispersed with two methods:
polymer adsorption of single-stranded DNA or miscelle
encapsulation with sodium deoxycholate surfactant.


DNA-wrapped CoMoCat SWCNTs (S-P95-02 Grade, Batch NI6-A001,
Southwest nanotechnologies) were prepared using (GT)$_{15}$
single-stranded DNA and separated by length using size exclusion
chromatography (SEC).\cite{zheng03,fagan07}  Sodium deoxycholate
(DOC) dispersions of SWCNTs from the same CoMoCat starting
material were prepared by substituting DOC ($2\,\%$ by mass) for
the DNA.  These DOC dispersions were length separated via
centrifugation.\cite{fagan08}  The resulting samples, which are
well-dispersed and length-separated fractions, exhibit
insignificant chirality enhancement.  After length separation,
these aqueous samples are then concentrated and dialyzed to remove
non-SWCNT components, apart from the dispersants.

\begin{figure}[hbt]
\begin{center}\leavevmode
\includegraphics[clip,width=3.25in]{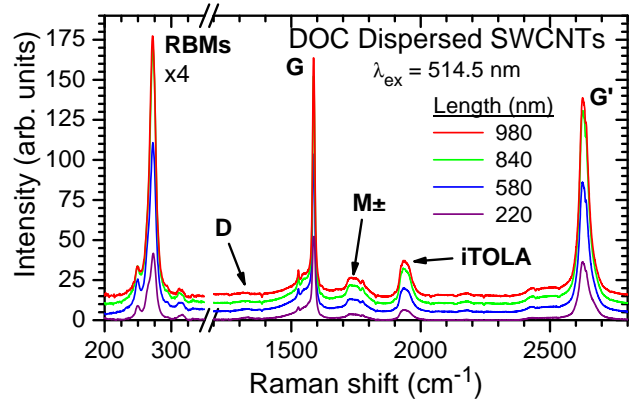}
\caption{Raman scattering from DOC dispersed SWCNT fractions with
laser excitation at 514.5$\,$nm, and scaled by the water
OH-stretch mode (not shown) for concentration. SWCNT phonon modes
are indicated. The D-band ($\approx 1328\,$\cm) is relatively
constant in length and substantially smaller than the RBM (scaled
$4\times$), G, and G$'$ peaks, all of which increase in intensity
with length.} \label{fig;RamanDOC}
\end{center}
\end{figure}

Raman spectra were collected from fractions ranging in length from
approximately $60\,$nm to larger than $1\,\mu$m. For optical
characterization by UV-vis transmission and NIR fluorescence, we
focus on two sets of length fractions, denoted $A$ and $B$, with
nearly identical lengths for the different dispersions of
approximately $L_A = 580\,$nm and $L_B = 230\,$nm.
Figure~\ref{fig;RamanDOC} shows a comparison of the Raman spectra
measured for different length fractions dispersed with DOC
surfactant.
The spectral intensities are scaled for SWCNT
concentration\cite{wu06} by the strength of the water OH-stretch
mode at $\approx 3400\,$\cm.  The scattering intensity (each
spectrum offset 5 units for clarity) increases monotonically with
the length of the fractions.  With laser excitation at
$514.5\,$nm, strong features due to the radial breathing modes
(RBMs), G, G$'$, iTOLA, and M$\pm$-bands\cite{dresselhaus05} are
clearly visible. The RBM spectral region shown in
Fig.~\ref{fig;RamanDOC} is scaled by a factor of four.

\begin{figure}[hbt]
\begin{center}\leavevmode
\includegraphics[width=3.25in]{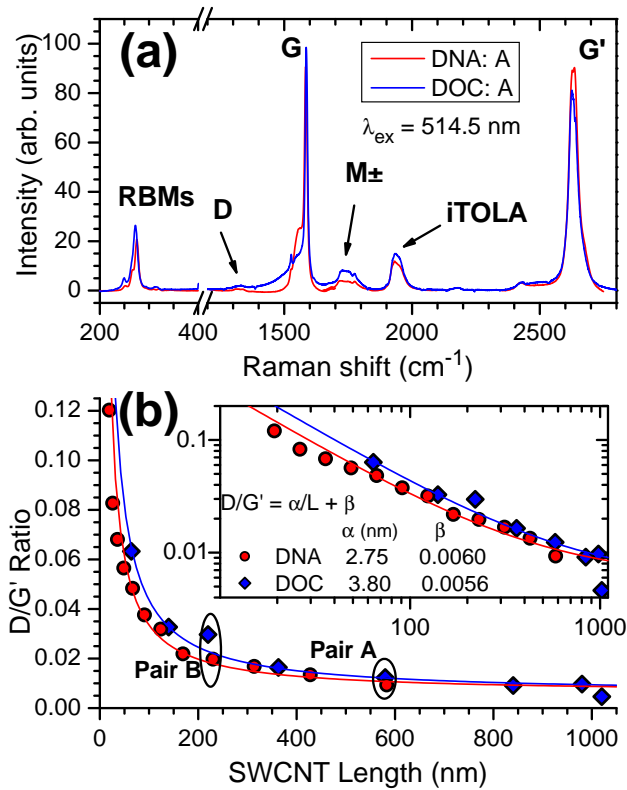}
\caption{(a) Raman spectra from the equivalent length $L_A \approx
580\,$nm, DOC and DNA dispersed SWCNT fractions with excitation at
$514.5\,$nm. The D-band has similar intensity for the two
dispersants.  (b) Length dependence of the D/G$'$\ ratio for the
DNA (red circles) and DOC (blue squares) dispersed SWCNTs. The
D/G$'$\ (and D/G) ratio for both decreases with increasing SWCNT
length. Equivalent length fraction pairs $A$ and $B$ are
indicated. Inset:  D/G$'$ ratio on a log-log scale.}
\label{fig;RamanDGp}
\end{center}
\end{figure}

Figure~\ref{fig;RamanDGp}(a) shows for comparison the Raman
spectra of the length fraction $L_A \approx 580\,$nm dispersed
using DOC (blue curve) and DNA (red curve). The D-band intensity
appears small compared to the G or G$'$\ modes and behaves
similarly for the two fractions. At $514.5\,$nm excitation, the
SWCNTs are off peak resonance ($E_{22}^S$ or $E_{11}^M$) for both
samples. Slight differences in the excitation energies of the two
dispersions result in an additional Breit-Wigner-Fano
contribution\cite{brown01} to the G-band from the partial
excitation of metallic SWCNTs in the DNA-wrapped sample.

The quantitative similarity for each of the D, G and G$'$\ bands
in the two samples shown in Fig.~\ref{fig;RamanDGp}(a) is
particularly striking.  The D-band mode requires in-plane
translational symmetry breaking to conserve momentum in a single
phonon scattering process.\cite{dresselhaus05} Disorder or
defects, including end cap effects, provide the necessary symmetry
breaking and give rise to the D-band. The second harmonic of the
D-band, G$'$, conserves momentum with two phonons and does not
require the presence of defects. The similarity of D and G
spectral features in the equivalent length SWCNTs implies a common
degree of disorder for the DNA and DOC dispersions.  The presence
of disorder/defects strongly affects the recombination of
excitons\cite{cognet07} and is thought to be one of the key
factors limiting the fluorescence quantum yield.

Thomsen and Reich\cite{thomsen07} suggest a comparison of the
D/G$'$\ intensity ratio as a metric more characteristic of defect
density than D/G.  Figure~\ref{fig;RamanDGp}(b) shows that the
ratio of the D-band ($\approx 1328\,$\cm) compared to G$'$-band
decreases with increasing SWCNT length.  DNA (red circles) and DOC
(blue squares) dispersants exhibit a common D/G$'$ ratio trend
over the entire range of length fractions. In both cases, the
D/G$'$\ ratio decays towards a small, constant value with an
approximately $1 / L$ length dependence. The D/G ratio (not shown)
behaves similarly.

Tuinstra and Koenig\cite{tuinstra70} originally reported the $1/L$
scaling behavior of the Raman D/G intensity with graphite
crystallite size. Can\c{c}ado \etal\cite{cancado06} observed a
similar $1/L$ scaling for graphite nanocrystals studied as a
function of excitation energy.  A similar $1/L$ dependence is
expected theoretically in SWCNTs for D-band contributions arising
from end caps.  Such behavior has been observed in short
($<100\,$nm) SWCNTs.\cite{chou07}  Thus, for defect-free SWCNTs,
the contribution from endcaps to the D/G$'$ ratio should scale
inversely with length, $\mathrm{D/G}' = \alpha/L$, where $\alpha$
is a length-independent constant. As shown in
Fig.~\ref{fig;RamanDGp}(b), our D/G$'$ ratio closely follows $1/L$
and then asymptotes to a small, constant value for the longest
fractions. Defect density along the nanotube axis should not scale
with length and hence provides a constant contribution to the
D/G$'$ ratio.  Fits of the D/G$'$ ratios to $\alpha/L + \beta$,
where $\beta$ is a constant defect contribution, are shown as
solid lines in Fig.~\ref{fig;RamanDGp}(b). The length at which the
$\alpha/L$ term and the constant term $\beta$ are equal,
represents a crossover between a dominant contribution from endcap
defects to those from tube axis defects. The length $L_0$ at the
crossover point, where $\alpha / L_0 = \beta$, may be thought of
as a characteristic length between defects. From the fits shown in
Fig.~\ref{fig;RamanDGp}(b), $L_0$ is approximately $(460 \pm
180)\,$nm for DNA and $(670 \pm 270)\,$nm for DOC. The defect
density $\eta$ is then given by $\eta = 1 / L_0$.

\begin{figure}[hbt]
\begin{center}\leavevmode
\includegraphics[clip,width=3.25in]{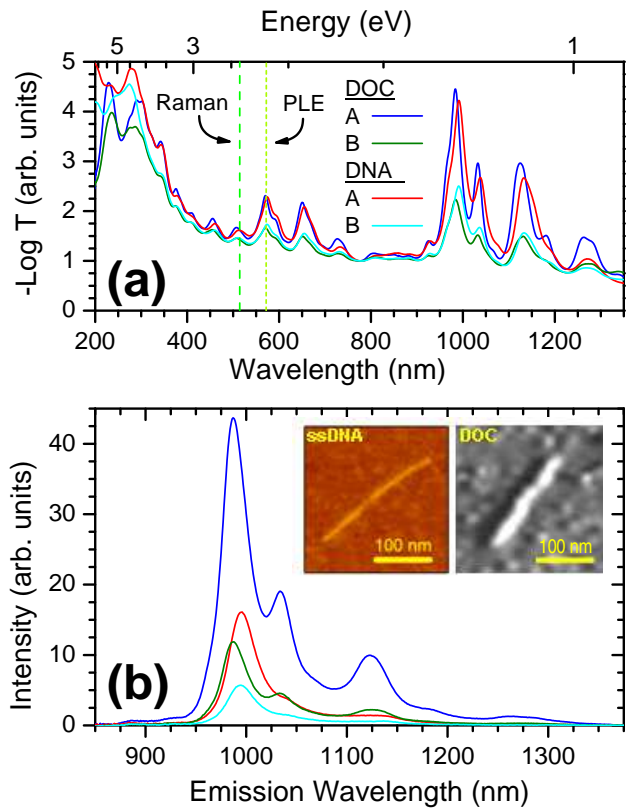}
\caption{(a) Log transmission spectra for two sets of SWCNT length
fractions with different dispersing agent:  ssDNA or DOC.  The
spectra have been scaled for concentration at $775\,$nm. Raman
excitation at $514.5\,$nm and fluorescence excitation at $572\,$nm
are indicated with dashed lines. (b) Fluorescence intensity
spectra for the samples shown in (a). Inset: Comparison of AFM
images for similar length SWCNTs wrapped with ssDNA and DOC. The
larger micellar structured formed by the DOC is visible in the
micrograph on the right.} \label{fig;AbsPL}
\end{center}
\end{figure}

The UV-vis transmission spectra of the two sets of length
fractions are shown in Fig.~\ref{fig;AbsPL}(a). Apart from a
slight red shift of the peak feature locations in the DNA
dispersions and the additional absorbance in the UV region
resulting from the bound DNA, the absorbance is nearly the same
for either dispersing agent.  However, as demonstrated in earlier
work,\cite{fagan07,fagan08} the spectral weight of the optical
transitions appears to depend on the nanotube length for
approximately equal SWCNT concentration. The peak features are
appreciably larger for the longer SWCNTs.
Figure~\ref{fig;AbsPL}(b) shows the fluorescent emission from the
two sets of samples diluted to normalize concentration as
determined by the absorption at $775\,$nm.  The fluorescence
exhibits a similar length dependence as the absorbance. Note that,
the SWCNT fluorescence is approximately $2 - 3$ times larger for
DOC versus DNA dispersed fractions of the same length. This
observation agrees with Haggenmueller \etal\cite{haggenmueller08}
who found DOC-dispersed SWCNTs display the strongest fluorescence
among roughly 20 dispersing agents for solutions containing a mix
of lengths. The Raman data reported here demonstrate comparable
defect density between the two dispersion methods, suggesting that
this difference in PL emission with dispersant is an environmental
effect.

\comment{The decreased emission intensity of DNA wrapped SWCNTs
can now be more strongly attributed to environmental effects as
the apparent defect density is similar for both sets of fractions.
}

%


Atomic force microscopy (AFM), shown in the inset of
Fig.~\ref{fig;AbsPL}(b), reveals a difference in the two surface
coatings. The natural tendency for the DOC to encapsulate the
SWCNT in a micelle leads to a larger overall diameter of the
entire hydrophyllic particle, with less area exposed to the
environment. Such a protective layer likely serves to preserve the
pristine character of the SWCNT graphitic lattice. In contrast,
the DNA physically adsorbs to the surface of the SWCNT in a
nonuniform manner.  Qian \etal\cite{qian08} found that non-uniform
DNA wrapping introduces a localized perturbation on the emission
of the SWCNTs, which likely accounts for the reduced PL.

\comment{
\begin{figure}[hbt]
\begin{center}\leavevmode
\includegraphics[clip,width=3.5in]{fig-AFM}
\caption{Comparison of AFM images for similar length SWCNTs
wrapped with ssDNA and DOC.  The micellar structured formed by the
DOC is likely to provide better isolation of the SWCNT from the
environment.} \label{fig;AFM}
\end{center}
\end{figure}
}


In conclusion, Raman scattering from SWCNTs dispersed with either
DNA or DOC depends on length, the D-band contribution results
primarily from end caps, and the defect density is approximately
equal for the two dispersants. Furthermore, the data suggest that
the inhomogeneity of polymer wrapping quenches fluorescent
emission from DNA-wrapped SWCNTs as compared to micellar-dispersed
SWCNTs. These results will have important implications in
biological applications that seek to exploit the intrinsic
fluorescent emission of SWCNTs.

\ \\
\bf{\it{Acknowledgments -}} J.R.S. acknowledges the support of a
NIST-National Research Council postdoctoral fellowship.



\begin{thebibliography}{99}

\bibitem{kitiyanan03} B.~Kitiyanan, W.~E.~Alvarez, J.~H.~Harwell, D.~E.~Resasco,
Chem. Phys. Lett. \bf{317}, 497 (2003).

\bibitem{zheng03} M.~Zheng, A.~Jagota, E.~D.~Semke, B.~A.~Diner, R.~S.~McLean, S.~R.~Lustig,
R.~E.~Richardson, N.~G.~Tassi, Nat. Mater. \bf{2}, 338 (2003).

\bibitem{fagan07} J.~A.~Fagan, J.~R.~Simpson, B.~J.~Bauer,
S.~Lacerda, M.~L.~Becker, J.~Chun, K.~B.~Migler,
A.~R.~Hight~Walker, E.~K.~Hobbie, J. Am. Chem. Soc. \bf{129},
10607 (2007).

\bibitem{fagan08} J.~A.~Fagan, M.~L.~Becker, J.~Chun,
E.~K.~Hobbie, Adv. Mat. \bf{20}, 1609 (2008).

\bibitem{wu06} Z.~Wu, C.~Zhang, P.~C.~Stair, Cat. Tod. \bf{113},
40 (2006).

\bibitem{dresselhaus05} M.~S.~Dresselhausa, G.~Dresselhaus, R.~Saito,
A.~Jorio, Phys. Rep. \bf{409}, 47 (2005).

\bibitem{brown01}  S.~D.~M.~Brown, A.~Jorio, P.~Corio,
M.~S.~Dresslehaus, G.~Dresselhaus, R.~Saito, and K.~Kneipp, Phys.
Rev. B \bf{63}, 155414 (2001).

\bibitem{cognet07}  L.~Cognet, D.~A.~Tsyboulski, J.~R.~Rocha,
C.~D.~Doyle, J.~M.~Tour, R.~B.~Weisman, Science \bf{316}, 1465
(2007).

\bibitem{thomsen07} C.~Thomsen and S.~Reich, Topics Appl. Physics
\bf{108}, 115 (2007).

\bibitem{tuinstra70}  F.~Tuinstra and J.~L.~Koenig, J. Chem. Phys. \bf{53}, 1126 (1970).

\bibitem{cancado06}  L.~G.~Con\c{c}ado \etal, App. Phys. Lett.
\bf{88}, 163106 (2006).

\bibitem{chou07} S.~G.~Chou, H.~Son, J.~Kong, A.~Jorio,
R.~Saito, M.~Zheng, G.~Dresselhaus, and M.~S.~Dresselhaus, Appl.
Phys. Lett. \bf{90}, 131109 (2007).

\bibitem{haggenmueller08} R.~Haggenmueller \etal,
Langmuir \bf{24}, 5070 (2008).

\bibitem{qian08} H.~Qian, P.~T.~Araujo, C.~Georgi, T.~Gokus, N.~Hartmann, A.~A.~Green,
A.~Jorio, M.~C.~Hersam, L.~Novotny, A.~Hartschuh, Nano Lett.
\bf{8}, 2706 (2008).


\end{thebibliography}


\end{document}